# Three-stage Acceleration of Solar Energetic Particles Detected by Parker Solar Probe

Xiaomin Chen[1,2] and Chuan Li[1,2,3]
[1] School of Astronomy and Space Science, Nanjing University, Nanjing 210023, People's Republic of China; lic@nju.edu.cn
[2] Key Laboratory of Modern Astronomy and Astrophysics (Nanjing University), Ministry of Education, Nanjing 210023, People's Republic of China
[3] Institute of Science and Technology for Deep Space Exploration, Suzhou Campus, Nanjing University, Suzhou 215163, People's Republic of China
Received 2024 February 26; revised 2024 April 29; accepted 2024 May 10; published 2024 May 28

## Abstract

Coronal mass ejections (CMEs) drive powerful shocks and thereby accelerate solar energetic particles (SEPs) as they propagate from the corona into interplanetary space. Here we present the processes of three-stage particle acceleration by a CME-driven shock detected by the in situ spacecraft—Parker Solar Probe (PSP) on 2022 August 27. The onset of SEPs is produced by a fast CME with a speed of 1284 km s$^{-1}$ when it propagates to ∼2.85 $R_\odot$. The second stage of particle acceleration occurs when the fast CME catches up and interacts with a preceding slow one in interplanetary space at ∼40 $R_\odot$ (∼0.19 au). The CME interaction is accompanied by an intense interplanetary type II radio enhancement. Such direct measurement of particle acceleration during interplanetary CME interaction/radio enhancement is rarely recorded in previous studies. The third stage of energetic storm particles is associated with the CME-driven shock passage of the PSP at ∼0.38 au. Obviously, harder particle spectra are found in the latter two stages than the first one, which can arise from a stronger shock produced by the CME interaction and the enriched seed particles inside the preceding CME.

*Unified Astronomy Thesaurus concepts:* Solar energetic particles (1491); Solar coronal mass ejections (310); Interplanetary shocks (829)

*Supporting material:* animation

## 1. Introduction

Solar energetic particles (SEPs) are produced by solar eruptions, notably flares and coronal mass ejections (CMEs), which have significant impacts on the Sun–Earth space environment, i.e., space-weather effects. The source regions and acceleration mechanisms of SEPs have been widely studied and continuously refined. Solar flares are usually responsible for impulsive SEP events with high electron and proton ratio, and $^3$He and $^4$He ratio, suggesting a stochastic acceleration during the process of magnetic reconnection (Reames 1999; Li et al. 2011). Such SEP events are frequently observed as the occurrence rate of solar energetic electron events reaches 10$^4$ yr$^{-1}$ at solar maximum (Wang et al. 2012). In large SEP events such as the ground-level enhancement events, CMEs are the main accelerators as they can drive shocks to accelerate particles over a large scale within the heliosphere (Reames 1999; Desai & Giacalone 2016). Based on the shock geometry and magnetic field topology, two types of acceleration mechanisms are proposed: diffusive shock acceleration (Axford et al. 1977; Zank et al. 2000; Kong et al. 2019) and shock drift acceleration (Holman & Pesses 1983; Guo & Giacalone 2010). The shock acceleration efficiency depends on both the shock properties and background plasma conditions such as the turbulence strength (Zimbardo et al. 2006; Guo et al. 2021) and seed particle richness (Tylka et al. 2005; Laming et al. 2013; Wijsen et al. 2023). However, figuring out the relation of particle production, shock property, and plasma condition remains challenging in observations.

To determine the source regions, the particle release times derived from the in situ measurement are always used to compare with the associated remote-sensing observations, such as the flare emission, radio bursts, and CME propagation (Reames 2009; Li et al. 2013; Wang & Qin 2015). During the CME-driven shock propagation from the corona into interplanetary space, two stages of particle acceleration are usually observed, including the particle release near the Sun and the energetic storm particles (ESPs) as the shock arrives (Gosling et al. 1981; Mäkelä et al. 2011). The interplanetary CME may interact with coronal streamers (Frassati et al. 2022), corotating interaction regions (Liu et al. 2019), and preceding CMEs (Shen et al. 2012; Lugaz et al. 2012, 2017). Notably, a preceding CME can be a source of particle injection, as it provides abundant seed particles and high turbulent environment (Li et al. 2012). A statistical study of Gopalswamy et al. (2002) had shown that a fast CME with a preceding one within ∼12 hours is more likely to be SEP-rich. The radio enhancement of interplanetary type II bursts may provide evidence for the process of CME interaction. Gopalswamy et al. (2001) proposed such radio enhancement as the shock strengthening when a fast CME-driven shock crossing the core of a preceding CME. Martínez Oliveros et al. (2012) used a direction-finding technique for type II analysis and confirmed that the radio enhancement is caused by the CME interaction. Such interplanetary CME interaction may produce another stage of particle acceleration, which, however, is rarely observed by in situ particle measurement.

In this study, we investigate the SEP event produced by a CME-driven shock during its propagation from the corona into interplanetary space on 2022 August 27. Three stages of particle acceleration was first recorded by the Parker Solar Probe (PSP, Fox et al. 2016), which was located at ∼0.38 AU when the event occurred. In Section 2, we introduce the instruments and their data applied in this study. Section 3 presents the remote-sensing solar observations and in situ particle measurements, as well as the analysis results. A brief summary and discussion are given in Section 4.

## 2. Instrumentation

The particle measurement of the SEP event was obtained from Integrated Science Investigation of the Sun (IS☉IS; McComas et al. 2016) on board the PSP. IS☉IS measures both the low-energy (EPI-Lo) and high-energy (EPI-Hi) bands of particles with Energetic Particle Instruments (EPI). EPI-Hi consists of three telescopes: High-energy Telescope (HET), Low-energy Telescope (LET) 1, and LET2. HET and LET1 detect particles from two directions of sunward and antisunward. LET2 detect particles from omnidirection. Here we use the data measured in the sunward direction. The data in the sunward direction for EPI-Lo are simply calculated by data averaged in the sunward detectors (L22, L25, L34–37, L44, and L46). The solar wind measurements including velocity, density, and temperature were obtained from the SPAN-Ai instrument of the SWEAP (Kasper et al. 2016) on board the PSP.

The associated solar eruption was observed by the Atmospheric Imaging Assembly (AIA; Lemen et al. 2012) on board the Solar Dynamics Observatory (SDO; Pesnell et al. 2012) and the Hα Imaging Spectrograph (HIS; Liu et al. 2022) on board the Chinese Hα Solar Explorer (CHASE; Li et al. 2022). The SDO/AIA provides extreme ultraviolet images of the corona. The CHASE/HIS provides spectral and imaging data of the chromosphere and photosphere. The CME white-light observations were obtained from the Large Angle and Spectrometric Coronagraph (LASCO; Brueckner et al. 1995) on board the Solar and Heliospheric Observatory (SOHO). The field of view (FOV) for LASCO/C2 and LASCO/C3 spans approximately 1.5 to 6 $R_\odot$ and 3.7 to 30 $R_\odot$, respectively. The interplanetary CME was observed by the Heliospheric Imager (HI; Howard et al. 2008) on board the STEREO A (Bougeret et al. 2008). The FOV of the HI-1 camera is 20°, centered 14° from the solar center. The corresponding radio emission was obtained from the Low-Frequency Receiver (LFR) and High-Frequency Receiver (HFR) of the WAVES instrument on board STEREO A. The LFR covers frequencies from 2.5 to 160 kHz, while the HFR covers frequencies from 125 kHz to 16.025 MHz.

## 3. Observations and Results

### 3.1. Event Overview

Figure 1 provides an overview of the SEP event that occurred on 2022 August 27. It was produced by a solar eruption from the southwest active region 13088. Panel (a) illustrates the relative positions of the spacecraft, whose data were used in this study. The viewpoint is from the north pole of the heliosphere. The Parker spiral of the interplanetary magnetic field (IMF) is calculated with an ideal solar wind model and found to be well connected to the solar source region and the PSP. Panel (b) shows the full-Sun image at Hα wave band observed by CHASE at 02:03 UT with the white box covering the solar active region, as well as the white-light image of the CME observed by LASCO/C2 at 02:35 UT. Three CME components are clearly shown as a bright core, a cavity, and a leading front with a diffusive shock sheath ahead. Panel (c) shows the proton intensity profiles at energy bands from 4 to 41 MeV. The vertical dashed lines indicate the three-stage particle injections, i.e., the onset of the SEP event at

∼02:23 UT, the second particle injection associated with the CME interaction started at ∼07:45–10:00 UT, and the ESPs during the shock passage from ∼14:08 to ∼18:00 UT.

### 3.2. Onset of the SEP Event

On 2022 August 27, the AR13088 located at S28W71 produced a solar eruption manifested as an M4.9 flare and a fast CME. Then, a high flux of energetic particles was detected by the in situ spacecraft PSP. By analyzing the imaging data of SDO/AIA, it is found that the solar eruption was triggered by the ejection of a magnetic flux rope (MFR). As shown in Panels (a)–(c) of Figure 2, a hot-channel-like MFR was observed in 94 Å images. To reveal the MFR evolution, we track the MFR along the yellow line in Panel (b). Panel (d) shows the height-time profile of the MFR evolution, which is fitted with a second-order polynomial function, as shown in Panel (e). The MFR underwent a slow rise phase and then was ejected with a nearly steady acceleration of $0.46 \pm 0.1$ km s$^{-2}$. The speed of the MFR reached ∼300 km s$^{-1}$ when it left the FOV of the AIA. The successful MFR eruption produced a typical three-component CME, which first appeared in the FOV of LASCO/C2 at 02:23 UT. We track the CME evolution through edge detection of the leading front in the image sequence. The height-time profile is displayed in Panel (c) of Figure 3. It is found that the CME expanded with a nearly uniform speed of $1284 \pm 27$ km s$^{-1}$ at 2.5–27 $R_\odot$.

We then derive the solar particle release (SPR) time of SEPs from the in situ measurement of the PSP. The top two panels of Figure 3 display the proton and electron flux observed by the EPI-Hi-LET detector. The proton flux shows a clear signature of velocity dispersion. We can derive the SPR time of protons by using the velocity dispersion analysis method: $T_{spr} = T_{onset} - L/v + 8.3$ min, where $T_{onset}$ is the onset time of particles at specific energy arriving at the in situ spacecraft, $v$ is the velocity of protons at different energies, and $L$ is the IMF path length connecting the source region and the spacecraft. To compare the SPR times with the solar remote-sensing observations by the near-Earth spacecraft, the light-travel time 8.3 minutes is taken into account. As shown by the black dashed line fitted in the top panel of Figure 3, the SPR time of protons is derived to be 02:23 UT ± 2 minutes, and the IMF path length is ∼0.6 au. For the relativistic electrons with energies ranging from 0.6 to 1.8 MeV, no clear velocity dispersion was observed as shown in the middle panel of Figure 3. We assume they propagate along the same IMF path length as the protons; their SPR time is then derived to be 02:23 UT ± 4 minutes, similar to the one of protons.

The derived SPR times are compared with the remote-sensing observations of the solar eruption, notably the flare and CME, as shown in the bottom panel of Figure 3. The soft X-ray flux of the M4.9 flare starts to rise at about 01:50 UT, reaches the peak at 02:40 UT, and lasts for about 3 hr. The deviation of the soft X-ray flux peaks at 02:15 UT, which is 8 minutes before the SPR times. Besides, there are no type III radio bursts observed at meter wavelength, suggesting that particles accelerated in the flare region do not escape from the low corona into interplanetary space. The height-time profiles of the MFR and the CME are overplotted as shown by colored asterisks. It is found that the release of SEPs takes place when the CME propagates to ∼2.85 $R_\odot$. Previous studies had also revealed that the effective CME-driven shock acceleration for large SEP events occurs when the CME propagates to 1.7–4.0 $R_\odot$ (Gopalswamy et al. 2012).

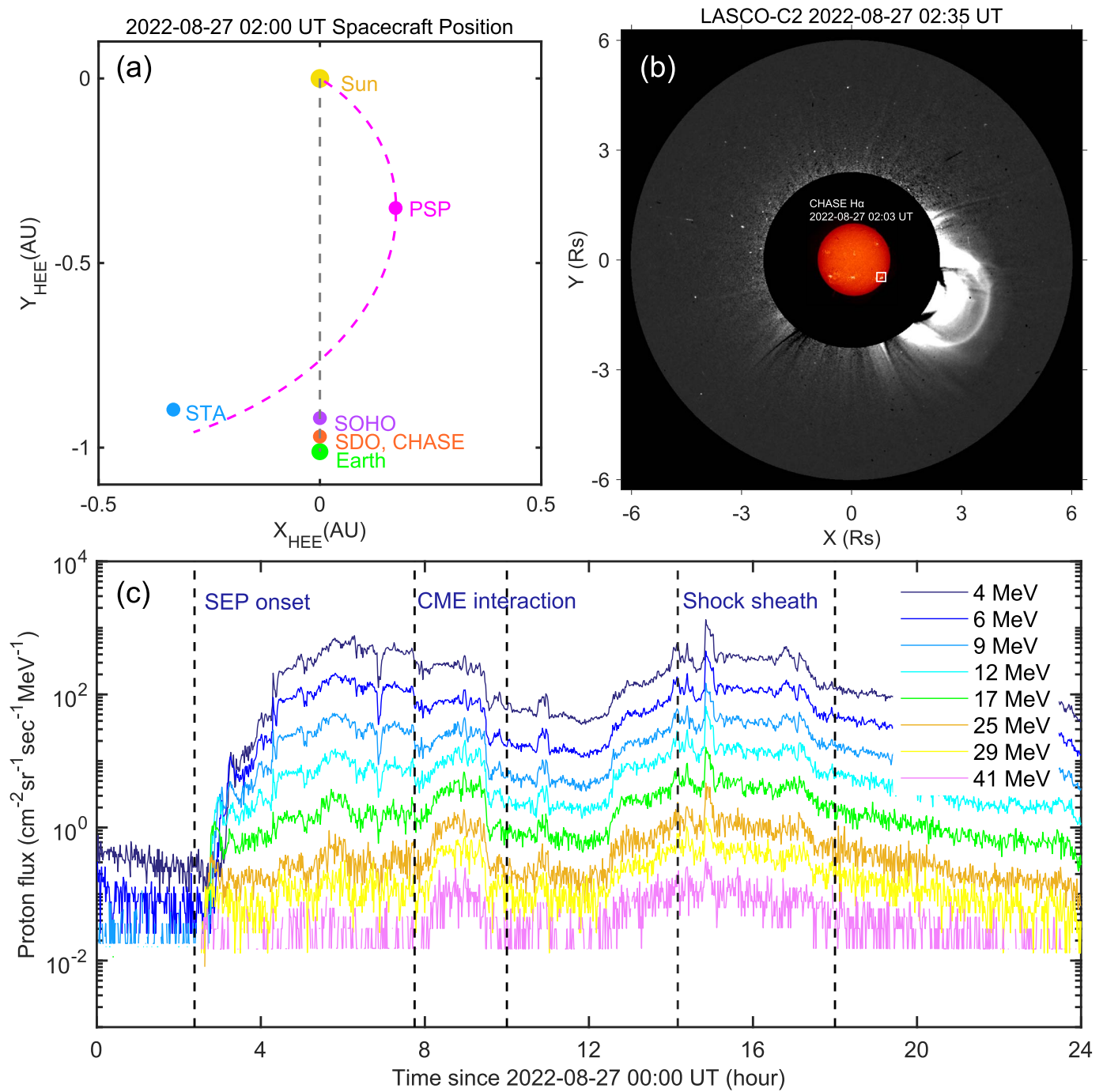

**Figure 1.** Overview of the SEP event on 2022 August 27. Panel (a) shows the positions of PSP (pink dot), STA (blue dot), SOHO (purple dot), and near-Earth spacecraft SDO and CHASE (orange dot). The pink dashed curve represents the Parker spiral of IMF. Panel (b) shows the Hα full-Sun image observed by CHASE and the white-light image of the CME observed by LASCO/C2. The white box marks the solar active region. Panel (c) shows the intensity-time profiles of protons observed by PSP. The three-stage particle injections are marked by the vertical dashed lines.

### 3.3. Particle Acceleration during the CME Interaction

The interplanetary CME propagation is captured by the HI-1 camera on board the STEREO A. We first process the white-light images in running-difference sequence and then apply a uniform filter to reduce the noise (see Nichitiu 2022). A preceding CME is found ahead of the primary one, as shown in Figure 4. According to the CDAW CME list,[4] the preceding CME occurred ∼12 hr earlier from the same active region, with a linear speed of ∼737 km s$^{-1}$. Panels (a)–(c) of Figure 4 present the process of the interplanetary CME interaction (see the Figure 4 animation). At 05:28 UT, the primary CME appeared in the FOV of the HI-1 camera, with a faint one in front of it. At 07:28 UT, the primary CME approached the preceding one, and the interaction might have started when its driven shock caught up the plasma in the preceding CME. At 14:08 UT, the two CMEs had merged into one and propagated farther in interplanetary space. Note that the exact time of CME interaction is not clear due to the poor temporal resolution (40 minutes) of the HI-1 instrument, which can be determined by the radio observation.

---

[4] https://cdaw.gsfc.nasa.gov/CME_list/

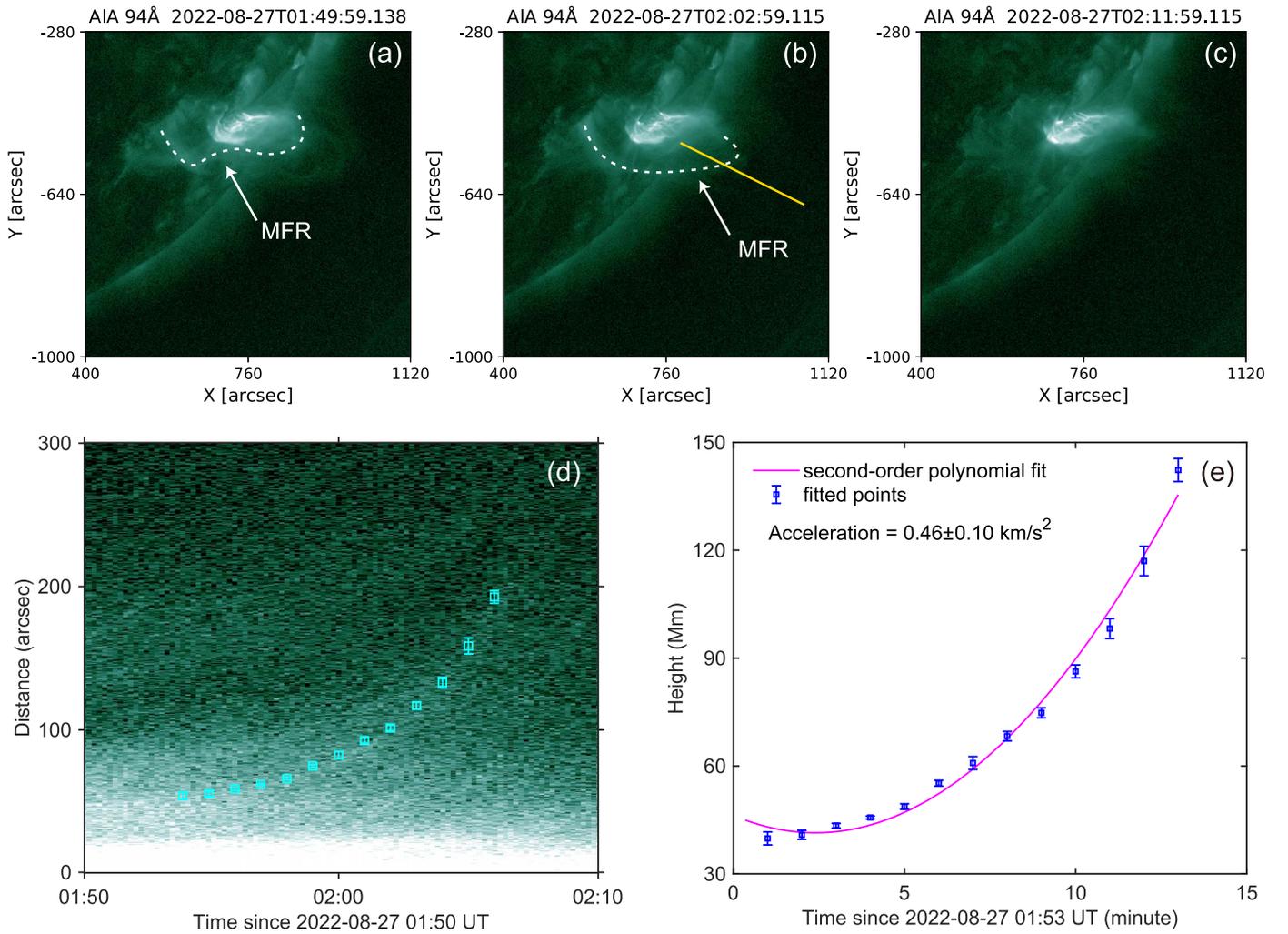

**Figure 2.** The ejection of the MFR observed at AIA 94 Å images. Panels (a)–(c) show the MFR before, during, and after ejection. The yellow line marks the direction of the MFR eruption, which is used for the time-slice analysis as shown in panel (d). Panel (e) shows the curve fitting of the height-time profile.

Panel (d) of Figure 4 presents the radio dynamic spectrum observed by the WAVES instrument on board the STEREO A. A long-lasting type II radio burst is marked by the white dashed curve. From ∼03:40 UT to ∼05:40 UT, scattered radio signals were observed with a frequency drift from 1.6 MHz to 375 kHz. An intense radio enhancement was observed from ∼07:45 to ∼10:00 UT, corresponding to the period of interplanetary CME interaction. It can be derived that the radio enhancement/CME interaction started when the primary CME reached ∼40 $R_\odot$. The radio enhancement shows low-frequency drift with a central frequency about 275 kHz. In a similar case studied by Ding et al. (2014), the type II enhancement was observed when the primary CME caught up with the trailing edge of the preceding one at ∼6 $R_\odot$, consistent with the time when electrons and protons were released. This event is probably the first one where the CME interaction and the accompanying radio enhancement have been directly examined with particle measurement at a distance of dozens of solar radii.

We next examine the proton flux shown in panel (c) of Figure 1. An obvious rise of proton flux is found during the period of radio enhancement/CME interaction. For protons at lower energies, i.e., 4 and 6 MeV, the background flux was relatively high and exhibited a declining trend, which leads to a plateau in the flux profile. For protons at higher energies, i.e., 9 to 41 MeV, the background flux is much lower; therefore, the proton flux during the interaction period shows a significant increase.

There is another possibility of giving rise to the second particle injection, i.e., other solar eruptions during this period. We have examined the multiwavelength solar observations and found two C-class flares both accompanied by narrow and slow CMEs. The two flares started at 07:47 UT and 09:11 UT, respectively. However, we found no concurrent rise of the proton flux measured by the EPI-Lo instrument on board the PSP, which covers energies ranging from ∼20 keV to 10 MeV. Panel (a) of Figure 5 shows a clear velocity dispersion from the SPR on the Sun, but no flux rise was observed during the period of radio enhancement. Therefore the rise of the proton flux during the period of radio enhancement is probably contributed by the ICME interaction.

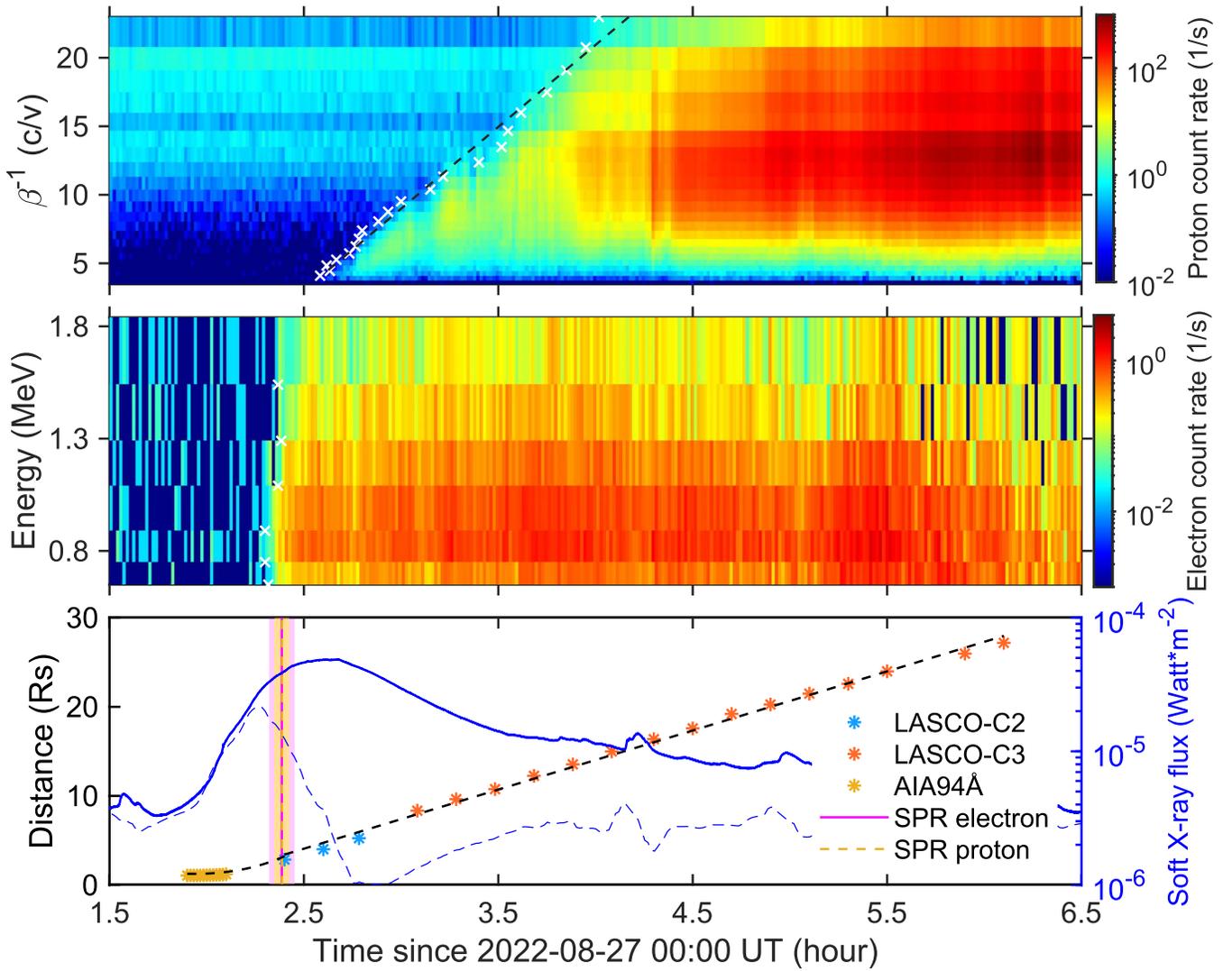

**Figure 3.** The proton (top panel) and electron (middle panel) flux detected by the EPI-Hi-LET detector on board the PSP. The overlaid black dashed line shows the fitting of the velocity dispersion of energetic protons. The white crosses mark the onset times of particles at specific energies recorded by the in situ spacecraft. The bottom panel shows the comparison of the SPR times and the solar eruption. The vertical shadow areas mark the SPR times of protons (pink) and electrons (yellow). The flare soft X-ray flux and its deviation are displayed by the blue solid and dashed curves, respectively. The height-time profiles of the MFR and the CME are shown by colored asterisks.

### 3.4. ESPs during the Shock Passage

We then focus on the phase of shock passage of the PSP at a heliocentric distance of 0.38 au. Figure 5 presents an overview of the ESP event. Panel (a) shows the spectrogram of protons measured by the EPI-Lo instrument with energies ranging from 20 keV to 10 MeV. Panel (b) shows the differential flux of protons at four energy bands. Panels (c)–(e) show the profiles of solar wind velocity, temperature, and density, which are measured by the SWEAP-SPAN instrument on board the PSP. Note that the magnetic field measurement was absent at the time. The first dashed line indicates the shock arriving and the second one indicates the passage of shock sheath, during which the ESPs were recorded. When the shock arrives, the velocity changes from 600 to 1068 km s$^{-1}$ in the upstream and downstream, the temperature from 70 to 437 eV, and the density from 40 to 104 cm$^{-3}$. Based on these parameters, one can derive the compression ratio of the CME-driven shock to be ~2.6.

As shown by the panel (a) of Figure 5, the proton flux exhibits a clear velocity dispersion from about 03:00 UT to 10:00 UT, which is produced by the first stage of acceleration near the Sun. As mentioned in Section 3.3, there were no obvious rises of proton fluxes at low energies during the CME interaction. During the shock's approach to the PSP, the proton fluxes, as shown by panel (b), increase slightly and experience a sudden enhancement during the shock passage at 14:08 UT. After that, the proton flux presents a slight decrease and reaches a plateau for ~1.5 hr. The proton flux decreases to the background levels until the shock sheath crossing at ~18:00 UT.

### 4. Summary and Discussion

In this study, we investigate the SEP event that occurred on 2022 August 27. For the first time, three stages of particle acceleration were recorded by the PSP located at a heliocentric distance of ~0.38 au. The onset of the SEP event is produced by a fast CME-driven shock when it propagates to ~2.85 $R_\odot$.

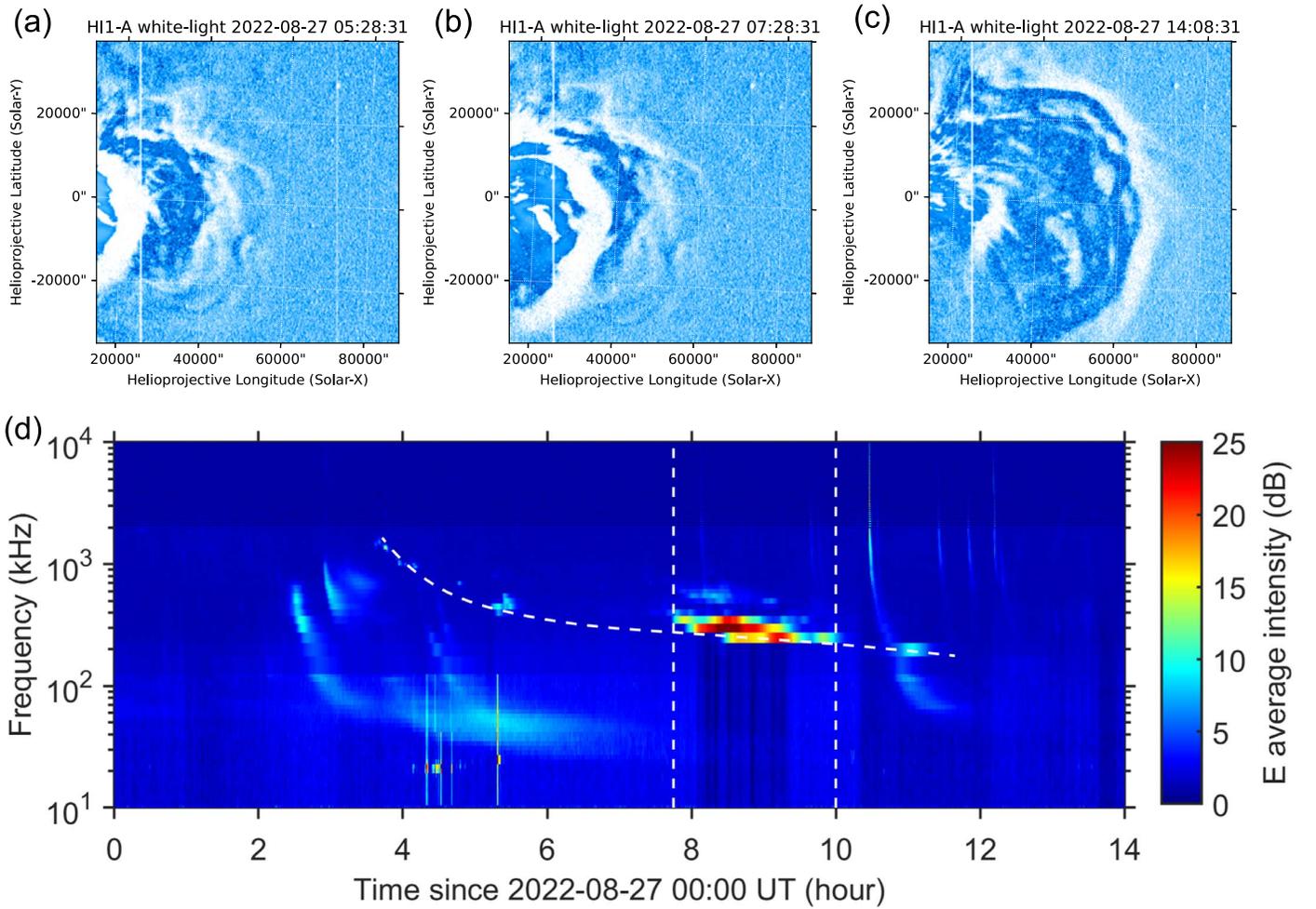

**Figure 4.** Observations of the interplanetary CME interaction. Panels (a)–(c) show the running-difference images observed by the HI-1 camera on board the STEREO A. Panel (d) shows the radio dynamic spectrum from the WAVES instrument of the STEREO A. The white dashed curve indicates a long-lasting type II radio burst. The vertical dashed lines indicate the type II radio enhancement during the CME interaction. The animation begins 2022 August 27 at 05:28:31 and ends the same day at 21:28:31. The real-time duration of the animation is 1.76 seconds.
(An animation of this figure is available.)

The second particle injection is associated with the CME interaction in interplanetary space, during which the primary CME catches up to a preceding slow one at ∼40 $R_\odot$ and gives rise to a remarkable enhancement of the type II radio burst. When the shock arrives at the in situ spacecraft, an obvious rise of particle flux, i.e., ESPs, is observed and lasts for ∼4 hr during the passage of the shock and the sheath region. This study provides clear evidence of particle acceleration of the CME-driven shock during its propagation from low corona to its passage at the in situ spacecraft.

To further clarify the acceleration mechanism, we have derived the proton spectra at the three stages of particle acceleration as shown in Figure 6. To align the particle measurement from the two telescopes of the EPI-Hi instrument, we have compared the data at the comparable energy bands (12–24 MeV) and divided the data from the LET telescope by a factor of ∼1.3 (Cohen et al. 2021). For the stage of particle release near the Sun, we integrate the proton differential intensities from 02:00 UT to 07:45 UT. The spectrum of the second stage is integrated from 07:45 UT to 10:00 UT, corresponding to the period of the CME interaction/radio enhancement. The spectrum of the ESP event is integrated from 14:08 UT to 18:00 UT, coinciding to the shock sheath passage.

All spectra have the background flux subtracted. For the particle flux measured by the EPI-HI instrument with energy ranging from 4 to 41 MeV, the spectra are fitted with the power-law function multiplied by an exponential turnover, referring to Ellison & Ramaty (1985), which is expressed as $dJ/dE \propto E^{-\gamma} exp(E/E_0)$, where $E_0$ is the turnover energy and $\gamma$ is the spectral index. The spectral indices of the three stages are derived to be 3.49 ± 0.36, 2.16 ± 0.56, and 2.59 ± 0.30, and the turnover energies are derived to be 25.10 ± 14.07 MeV, 17.51 ± 10.66 MeV, and 16.90 ± 5.32 MeV, respectively. In the third stage, for the proton flux with energy ranging from 86 to 489 keV, the spectrum is fitted with a single power-law function. The spectral index is then derived to be 2.44 ± 0.05, which is comparable to the one of 4–41 MeV. It is found that the latter two spectra present smaller spectral indices and smaller turnover energies compared to the first one.

Based on the diffusive shock acceleration, the particle spectral index is directly related to the shock compression ratio, i.e., the shock strength. Therefore, the latter two-stage harder spectra probably arose from a stronger shock formed during the CME interaction. A case study of Gopalswamy et al. (2001) suggested that the shock compression ratio could be enhanced during the CME interaction due to the decrease of the background Alfvén velocity within the preceding CME. Furthermore, a simulation of the CME interaction by Lugaz et al. (2005) indicates that the shocks driven by two CMEs can merge into one stronger shock, leading to a larger shock compression ratio.

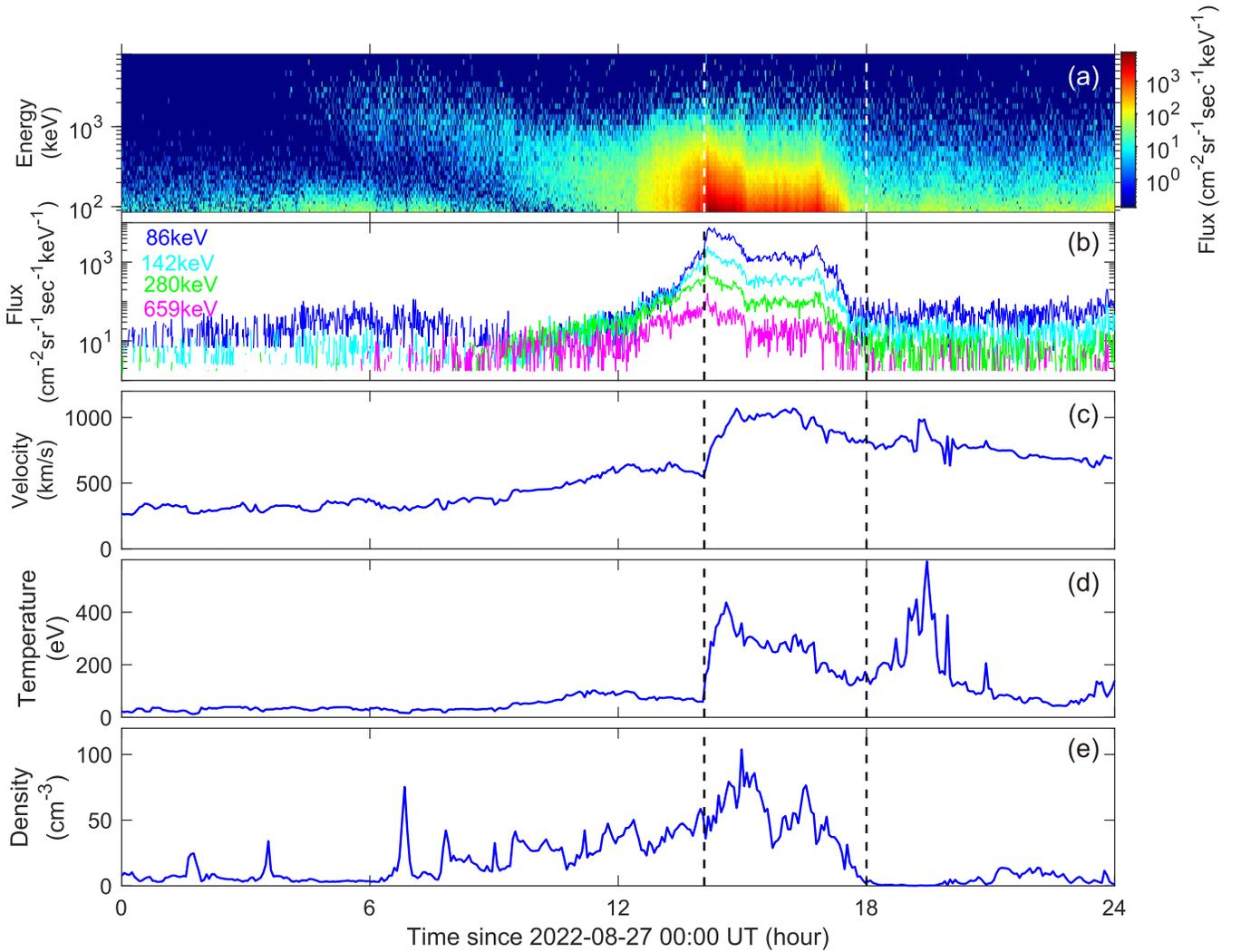

**Figure 5.** ESPs during the shock passage. Panel (a) shows the spectrogram of proton fluxes measured by the EPI-Lo instrument. The proton differential intensities at four energy bands are shown in panel (b). Panels (c)–(e) show the solar wind parameters of velocity, temperature, and density measured by the SWEAP-SPAN instrument. The two vertical dashed lines indicate the passage of the shock and the sheath region.

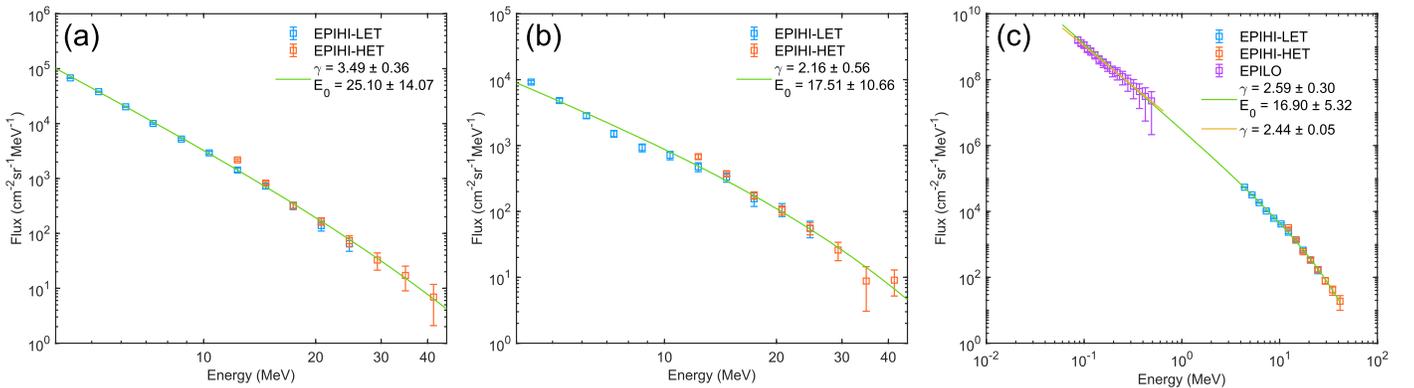

**Figure 6.** Proton spectra of the three stages of particle acceleration. The proton fluxes are obtained from the EPI-Hi and EPI-Lo detectors of IS☉IS on board the PSP.

The exponential turnover deviating from a single power-law spectrum arises from a variety of effects, such as the adiabatic deceleration and the finite size and lifetime of the CME-driven shock. These effects all produce spectra that turn over at higher energies when the diffusion coefficients increase with energies (Ellison & Ramaty 1985). Therefore, the turnover energies are related to the level of turbulence in front of the CME-driven shock. In this event, the turnover energies decrease at the three stages of particle acceleration, suggesting the CME interaction does not enhance the turbulence level significantly. Strong turbulence near the Sun might give rise to larger turnover energy in the first stage. Such strong turbulence occurs in the downstream of a preceding CME, which increases the scattering in the upstream of the primary CME-driven shock (Li et al. 2012).

During the interplanetary propagation of the CME-driven shock, the suprathermal particles in the solar wind provide a crucial part of the seed population. In particular, the energetic particles inside a preceding CME are expected to be more abundant and have a broader energy range. Therefore, it is likely easier for those particles to be reaccelerated by the primary CME-driven shock to reach higher energies, leading to a harder particle spectrum (Gopalswamy et al. 2004). As shown in the panel (c) of Figure 1, the proton intensity profiles at the higher energies, e.g., 41 MeV, show significant enhancements during the CME interaction compared to the first stage of particle release near the Sun.

The interplanetary transport may also influence the in situ particle spectrum. Particles with lower energies have a smaller mean free path and therefore experience stronger scattering during their propagation from the release site to the in situ spacecraft. This produces a harder particle spectrum. Such spectral hardening had been observed in large, gradual SEP events (Tylka et al. 2000) and in energetic particles produced by the corotating interaction regions (Zhao et al. 2016). In this event, however, harder spectra are found in the last two stages, indicating the CME interaction rather than the transport effect is responsible for the difference of particle spectra.

## Acknowledgments

We are grateful to the PSP, SDO, STEREO, SOHO, and CHASE teams for their invaluable support in providing observational data. We thank the anonymous referee whose comments and suggestions greatly improved this study. This study is supported by the NSFC under grant 12333009 and the CNSA project D050101.

## ORCID iDs

Xiaomin Chen  https://orcid.org/0009-0003-3760-705X
Chuan Li  https://orcid.org/0000-0001-7693-4908

## References

Axford, W. I., Leer, E., & Skadron, G. 1977, ICRC, 11, 132, (Budapest)
Bougeret, J. L., Goetz, K., Kaiser, M. L., et al. 2008, SSRv, 136, 487
Brueckner, G. E., Howard, R. A., Koomen, M. J., et al. 1995, SoPh, 162, 357
Cohen, C. M. S., Christian, E. R., Cummings, A. C., et al. 2021, A&A, 656, A29
Desai, M., & Giacalone, J. 2016, LRSP, 13, 3
Ding, L.-G., Li, G., Jiang, Y., et al. 2014, ApJL, 793, L35
Ellison, D. C., & Ramaty, R. 1985, ApJ, 298, 400
Fox, N. J., Velli, M. C., Bale, S. D., et al. 2016, SSRv, 204, 7
Frassati, F., Laurenza, M., Bemporad, A., et al. 2022, ApJ, 926, 227
Gopalswamy, N., Xie, H., Yashiro, S., et al. 2012, SSRv, 171, 23
Gopalswamy, N., Yashiro, S., Kaiser, M. L., Howard, R. A., & Bougeret, J. L. 2001, ApJL, 548, L91
Gopalswamy, N., Yashiro, S., Krucker, S., Stenborg, G., & Howard, R. A. 2004, JGRA, 109, A12105
Gopalswamy, N., Yashiro, S., Michałek, G., et al. 2002, ApJL, 572, L103
Gosling, J. T., Asbridge, J. R., Bame, S. J., et al. 1981, JGR, 86, 547
Guo, F., & Giacalone, J. 2010, ApJ, 715, 406
Guo, F., Giacalone, J., & Zhao, L. 2021, FrASS, 8, 27
Holman, G. D., & Pesses, M. E. 1983, ApJ, 267, 837
Howard, R. A., Moses, J. D., Vourlidas, A., et al. 2008, SSRv, 136, 67
Kasper, J. C., Abiad, R., Austin, G., et al. 2016, SSRv, 204, 131
Kong, X., Guo, F., Chen, Y., & Giacalone, J. 2019, ApJ, 883, 49
Laming, J. M., Moses, J. D., Ko, Y.-K., et al. 2013, ApJ, 770, 73
Lemen, J. R., Title, A. M., Akin, D. J., et al. 2012, SoPh, 275, 17
Li, C., Firoz, K. A., Sun, L. P., & Miroshnichenko, L. I. 2013, ApJ, 770, 34
Li, C., Matthews, S. A., van Driel-Gesztelyi, L., Sun, J., & Owen, C. J. 2011, ApJ, 735, 43
Li, C., Fang, C., Li, Z., et al. 2022, SCPMA, 65, 289602
Li, G., Moore, R., Mewaldt, R. A., Zhao, L., & Labrador, A. W. 2012, SSRv, 171, 141
Liu, Q., Tao, H., Chen, C., et al. 2022, SCPMA, 65, 289605
Liu, Y., Shen, F., & Yang, Y. 2019, ApJ, 887, 150
Lugaz, N., Farrugia, C. J., Davies, J. A., et al. 2012, ApJ, 759, 68
Lugaz, N., Manchester, W. B. I., & Gombosi, T. I. 2005, ApJ, 634, 651
Lugaz, N., Temmer, M., Wang, Y., & Farrugia, C. J. 2017, SoPh, 292, 64
Mäkelä, P., Gopalswamy, N., Akiyama, S., Xie, H., & Yashiro, S. 2011, JGRA, 116, A08101
Martínez Oliveros, J. C., Raftery, C. L., Bain, H. M., et al. 2012, ApJ, 748, 66
McComas, D. J., Alexander, N., Angold, N., et al. 2016, SSRv, 204, 187
Nichitiu, M. D. 2022, OJAp, 5, 8
Pesnell, W. D., Thompson, B. J., & Chamberlin, P. C. 2012, SoPh, 275, 3
Reames, D. V. 1999, SSRv, 90, 413
Reames, D. V. 2009, ApJ, 693, 812
Shen, C., Wang, Y., Wang, S., et al. 2012, NatPh, 8, 923
Tylka, A. J., Boberg, P. R., McGuire, R. E., Ng, C. K., & Reames, D. V. 2000, in AIP Conf. Proc. 528, Acceleration and Transport of Energetic Particles Observed in the Heliosphere, ed. R. A. Mewaldt et al. (Melville, NY: AIP), 147
Tylka, A. J., Cohen, C. M. S., Dietrich, W. F., et al. 2005, ApJ, 625, 474
Wang, L., Lin, R. P., Krucker, S., & Mason, G. M. 2012, ApJ, 759, 69
Wang, Y., & Qin, G. 2015, ApJ, 799, 111
Wijsen, N., Li, G., Ding, Z., et al. 2023, JGRA, 128, e2022JA031203
Zank, G. P., Rice, W. K. M., & Wu, C. C. 2000, JGR, 105, 25079
Zhao, L., Li, G., Ebert, R. W., et al. 2016, JGRA, 121, 77
Zimbardo, G., Pommois, P., & Veltri, P. 2006, ApJL, 639, L91